\documentstyle[aps]{revtex}
\begin{document}

\title{Comment on ``Quantum mechanics of smeared particles"}
\author{F. Brau\thanks{FNRS Postdoctoral Researcher; E-mail : fabian.brau@umh.ac.be}}
\address{Service de Physique G\'en\'erale et de Physique des Particules El\'ementaires, Groupe de Physique 
Nucl\'eaire Th\'eorique, Universit\'e de Mons-Hainaut, Mons, Belgique}
\date{\today}

\maketitle

\begin{abstract}
In a recent article, Sastry has proposed a quantum mechanics of smeared particles. 
We show that the effects induced by the modification of the Heisenberg algebra, proposed to take into 
account the delocalization of a particle defined via its Compton wavelength, are important enough to be 
excluded experimentally.
\end{abstract}

The idea to represent a particle not as an idealized point particle but instead as a
{\it smeared} particle is not new, but recently a subtle way to introduce this smearing has been
proposed. This formalism, used in Ref. \cite{sast00} with some modifications, was first developed 
by Kempf {\it et al.} \cite{kemp}. The idea is to modify the commutation relations between position 
and momentum (the Heisenberg algebra) to
introduce a new short-distance structure characterized by a finite minimal uncertainty
$\Delta x_0$ in position measurements. The existence of this minimal observable length is 
suggested by quantum gravity and string theory~\cite{gros88,amat89,magg93,amel97,haro98}.
In this context, the new short distance behaviour would arise at the Planck scale, and $\Delta x_0$ 
would correspond to a fundamental quantity closely linked with the structure of space-time \cite{kemp98}.
Kempf has suggested that this formalism could also be used to describe, as an effective theory, non-pointlike 
particles like hadrons, quasi-particles or collective excitations~\cite{kemp97b}.
In this case, $\Delta x_0$ is interpreted as a parameter linked with the structure of these particles and 
their finite size; no attempt is performed to give an explicit link between this parameter and some 
fundamental property of the particle: it is considered as a free parameter. 

In a recent article, Sastry suggests that the deformation parameter of
the Heisenberg algebra is given by the Compton wavelength of the particle \cite{sast00}. 
He points out that in the case of the hydrogen atom, and in general in
the quantum theory of atoms, the quantum mechanics of point particles gives an accurate
description because the characteristic size of the smearing of the electron (the Compton
wavelength) is $\alpha$ (the fine structure constant) times smaller than the characteristic
size of the atom $a_0$ (the Bohr radius). Even if this assertion is correct the effects
of this smearing of the electron are still too large and can be 
excluded by comparison between {\it standard} theoretical calculation and experimental data. 

The modification of the energy level positions of the hydrogen atom introduced
by the use of the new commutation relations between position and momentum
has been evaluated to first order in Ref. \cite{brau99}. 
The order of magnitude of the correction is given by $(\Delta x_0)^2\, m^3\,
\alpha^4$, where $m$ is the mass of the electron ($\hbar=c=1$). The use of the Compton wavelength
as the deformation parameter of the Heisenberg algebra, $\Delta x_0 \propto 1/m$, leads 
to a correction of the same order ($10^{-3}$ eV) as the first relativistic kinematic and 
the spin-orbit corrections 
(which describe the fine structure of the hydrogen energy levels) and is thus
two orders of magnitude larger than the Lamb shift and hyperfine structure 
corrections (see for example \cite{bjor}). The agreement between {\it standard} theory
and experiment is about 1 MHz ($10^{-8}$ eV) for the Lamb Shift of the 1S state 
\cite{hell,kars02} and about 0.1 MHz for the hyperfine structure of the
1S state (the famous 21cm hyperfine transition)\cite{udem97,mall98}. This excellent agreement
definitely excludes the proposal of Sastry.

\end{document}